\documentclass[conference]{IEEEtran}
\usepackage{milcom}

\usepackage[inline]{enumitem}

\title{NAC: Automating Access Control via Named Data
	\thanks{This work is partially supported by US National Science Foundation under award CNS-1719403.}
}

\author{
\IEEEauthorblockN{Zhiyi Zhang}
\IEEEauthorblockA{
\textit{UCLA}\\
zhiyi@cs.ucla.edu}
\and
\IEEEauthorblockN{Yingdi Yu}
\IEEEauthorblockA{
\textit{UCLA}\\
yingdi@cs.ucla.edu}
\and
\IEEEauthorblockN{Sanjeev Kaushik Ramani}
\IEEEauthorblockA{
\textit{Florida Int'l University}\\
skaus004@fiu.edu}
\and
\IEEEauthorblockN{Alex Afanasyev}
\IEEEauthorblockA{
\textit{Florida Int'l University}\\
aa@cs.fiu.edu}
\and
\IEEEauthorblockN{Lixia Zhang}
\IEEEauthorblockA{
\textit{UCLA}\\
lixia@cs.ucla.edu}
}

\begin{document}

\maketitle

\begin{abstract}
In this paper we present the design of Name-based Access Control (NAC) scheme, which supports data confidentiality and access control in Named Data Networking (NDN) architecture by encrypting content at the time of production, and by automating the distribution of encryption and decryption keys.
NAC achieves the above design goals by leveraging specially crafted NDN naming conventions to define and enforce access control policies, and to automate the cryptographic key management.
The paper also explains how NDN's hierarchically structured namespace allows NAC to support fine-grained access control policies, and how NDN's Interest-Data exchange can help NAC to function in case of intermittent connectivity.
Moreover, we show that NAC design can be further extended to support Attribute-based Encryption (ABE), which supports access control with additional levels of flexibility and scalability.

\end{abstract}


\section{Introduction}
\label{sec:intro}

Named Data Networking (NDN)~\cite{ndn}, a proposed Internet architecture, enables applications to retrieve desired data by names at the network layer. This is a fundamental departure from IP networking where one retrieves data by using the addresses of data containers. 
In addition to enabling efficient and robust data dissemination, NDN also introduces a data-centric security model by securing data directly~\cite{securityTR}, enabling end-to-end security regardless of the security, or lack of it, of communication channels and any other intermediaries. 

A common requirement of distributed applications is an effective and usable access control solution to ensure that only authorized users and applications can have access to certain contents.
Numerous access control approaches~\cite{kumar2006medium, RFC4949, goyal2006attribute,cp-abe,jahid2011easier} have been proposed; however, implementing these solutions over TCP/IP protocol stack requires non-trivial and error-prone configurations at the network layer for content retrieval and access key distribution.
Furthermore, utilizing third-party services like DNS for key storage and distribution also increases the attack surface of the overall system.

This paper describes the Name-based Access Control (NAC) scheme, which provides content confidentiality and access control in an NDN network.
NAC is built on a combination of symmetric and asymmetric cryptography algorithms, and utilizes NDN's data-centric security and naming convention to automate data access control.
Throughout the paper, we show that NAC scheme has following desirable properties:
\begin{enumerate} [label=(\roman*)]
	\item NAC leverages a specially crafted NDN naming convention to name cryptographic keys, enabling them to be retrieved automatically.
	\item NAC supports fine-grained access control through simple namespace design.
	\item NAC utilizes NDN's stateful forwarding plane and in-network storage to enable resilient communications in face of intermittent connectivity.
\end{enumerate}

We describe two implementations~\cite{nac-rsa, nac-abe} of the NAC scheme called NAC and NAC-ABE.
The two share the key properties, but differ in the asymmetric encryption algorithms used: the former uses RSA and the latter Ciphertext-Policy Attribute-based Encryption (CP-ABE)~\cite{cp-abe}.
Utilizing Attribute-based Encryption, NAC-ABE supports data access control with additional levels of flexibility and scalability.




With the assumption that readers are familiar with the basic concepts of NDN, which are also described in a companion paper~\cite{milcom18-ndn-overview}, we organize this paper as follows.
We describe, in Section~\ref{sec:nac-overview}, a simple battlefield application scenario, and introduce the assumptions and goals of NAC, together with a brief explanation of how NAC works in Section~\ref{sec:model}.
We introduce our implementation of NAC-ABE in Section~\ref{sec:nac-abe}.
In Section~\ref{sec:properties}, we explain how NAC scheme provides automatic key distribution and fine-grained access control, and how NDN enables NAC to operate with intermittent connectivity.
We evaluate the security and performance of the NAC design in Sections~\ref{sec:assessment} and \ref{sec:evaluation}.
We compare the access control system over TCP/IP and NDN, discuss the open issues of our design in Section~\ref{sec:disussion}, and conclude the work in Section~\ref{sec:conclusion}.

\section{Example Scenario}
\label{sec:model}

\begin{figure}[htbp]
	\centering
	\includegraphics[scale=0.6]{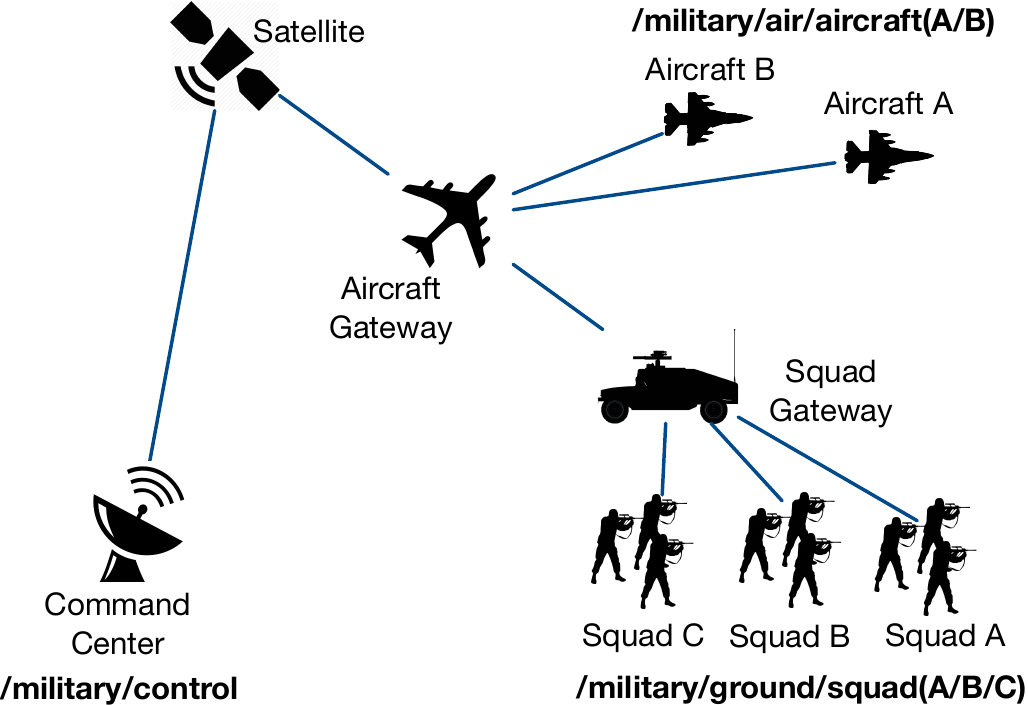}
	\caption{A battlefield communication scenario}
	\label{fig:model}
\end{figure}

To facilitate explanation and discussion in the rest of this paper, we first introduce a typical battlefield application scenario (see Fig.~\ref{fig:model}).
In a modern battlefield, there could be multiple types of entities that need to work in unison and coordinate together through an intermittent network with relatively high packet loss rate. 
One of the growing requirements of such military communications is strong confidentiality and effective access control. 

As shown in the Fig.~\ref{fig:model}, there are three types of entities in our network scenario:
\begin{enumerate*} [label=(\roman*)]
\item \emph{Command Center} (\name{/military/control}) that determines the access privileges of participants in a system.

\item \emph{Units} (\name{/military/air/aircraft(A/B)]} and \name{/military/ground/squad(A/B/C)}) that communicate with the command center and other units. For example, an aircraft may receive commands from the command center and provide surveillance information to a squad.

\item \emph{Forwarders} (Satellite, Aircraft Gateway, and Squad Gateway) that forward packets among units and command center. For example, satellites, aircraft gateways, and squad gateways may form a broadcast connectivity, and aid in forwarding packets.
\end{enumerate*}

This battlefield scenario requires that only authorized entities can access certain pieces of data. 
As an example, when the command center sends a command intended to only squad A, only soldiers from squad A could be able to read the content; all the other entities such as aircraft A and squad B should not be able to see the content even if they have retrieved the Data packets. 
In the rest of the paper, we will use this scenario to illustrate how NAC aids in providing effective communication confidentiality and effective access control.

\section{Name-based Access Control}
\label{sec:nac-overview}

\subsection{Assumptions and Goals}

The design of NAC assumes that proper trust relationships among entities in the system have already been established.
To be more specific, 
\begin{enumerate*} [label=(\roman*)]
\item each entity in the system has its own public/private key pair, 
and 
\item each entity is able to authenticate the Data packets produced by others through digital signature validation.
\end{enumerate*}
For example, a squad is able to authenticate the Data packet received from the command center.
This could be realized by security bootstrapping process as described in \cite{securityTR}.

Unlike traditional network-layer access control that focuses on the access to medium, NAC aims to control the access to the content of Data packets with several additional goals:
\begin{enumerate*} [label=(\roman*)]
\item access control can be done at fine granularities;
\item enforcement of the access control is automated as much as possible;
\item the system is robust against the intermittent network connectivity.
\end{enumerate*}

\subsection{Design Overview}

\begin{figure}[htbp]
	\centering
	\includegraphics[scale=0.4]{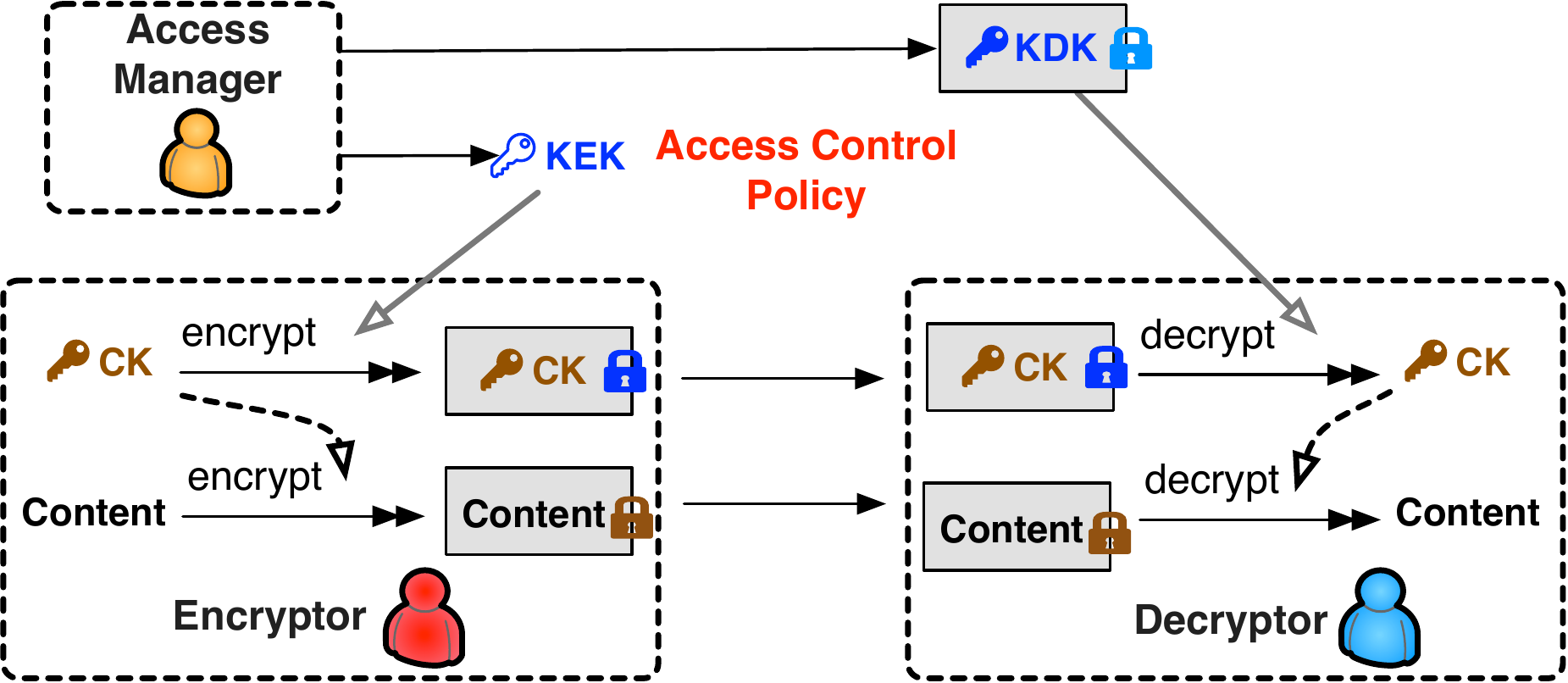}
	\caption{NAC Scheme}
	\label{fig:nac-model}
\end{figure}

NAC achieves aforementioned goals by using a combination of symmetric and asymmetric keys (see Fig.~\ref{fig:nac-model}) and utilizing NDN's structured, semantically meaningful naming to express the access policy and granularity.
In NAC design, there is an \emph{access manager} (e.g., command center), who defines the access control policies in a given system.
The access manager publishes its access control policies as a list of named public and private key pairs, called KEK (key-encryption key, public key) and KDK (key-decryption key, private key).
Leveraging the naming convention, a KEK's name indicates the \emph{granularity} (i.e., content name prefix) under which the Data packets should be encrypted with this KEK. 
On the other hand, the KDK name encodes both the name of the authorized granularity and name of the consumer to whom the access is granted.
To control the access rights, the access manager distributes KDK (key-decryption key) to authorized \emph{decryptors} by publishing KDK Data packets encrypted using decryptors' public keys.
\emph{Encryptors} are entities that publish encrypted content and they retrieve the named KEKs as the access control policy.
This named policy can be configured or inferred from configuration and data name (see Section~\ref{sec:auto} for an example of it).
Content is not directly encrypted using the KEK, but a symmetric content key, called the Content Key (CK); CK will then be encrypted using the KEK.
As a result, an encryption (or decryption) key chain can be established from an encryptor to a decryptor under the control of the access manager.

In NAC, the access manager, encryptor, and decryptor are different in terms of their function.
In practice, they can be a single entity based on specific application scenarios.
For instance, in our battlefield scenario, the command center is the access manager, encryptor (when sending commands to others), and decryptor (for received responses).

NAC achieves fine-grained access control by requiring the encryptor to follow the KEK name to encrypt content.
The more specific the KEK name is, the fewer Data packets can be decrypted using the corresponding KDK.
NAC also allows access manager to control the access through KDK distribution.
The more KDKs a decryptor can obtain, the more Data packets it can access.

NAC achieves automation of access control by publishing access control policy (named KEKs) and decryption keys (CKs and KDKs) as normal NDN Data packets.
As long as the named KEKs have been published in the network, encryptors can automatically retrieve them by name to encrypt content/CK properly.
The access manager and encryptors also publish the KDKs/CKs respectively, so that decryptors can follow the key names encoded in the encrypted Data packets to retrieve decryption keys and construct the decryption key chain automatically.

Since all the keys are just normal NDN Data packets, the data-centric security model of NDN allows the access manager and encryptors to freely distribute these Data packets within the network even to non-trusted data repository or in-network caches.
NDN's name-based retrieval ensures that as long as there is a copy for those keys in the network, encryptors and decryptors can always retrieve them even with intermittent connectivity.

\subsection{NAC Naming Conventions}
\label{sec:naming}

All the keys in NAC are named under a specific naming convention.
As further discussed in Section~\ref{sec:properties}, NAC leverages naming conventions to realize automatic key distribution and fine granularity of data confidentiality and access control.

\subsubsection{Key Encryption Key (KEK)}

Encryptors need to fetch KEK to encrypt the content.
NAC defines the naming convention for KEK and KEK Data packet.
\begin{center}
	KEK Name = \name{/<access manager prefix>/NAC/<granularity>/KEK/<key-id>}
\end{center}
where the access manager indicates the producer of the KEK, the granularity is the name prefix of the data that is being produced by the encryptor, and the key-id is the unique identifier of the key.

\subsubsection{Key Decryption Key (KDK)}

In NAC, the KDK follows the same convention as KEK except the name component \name{KEK}:
\begin{center}
	KDK Name = \name{/<access manager prefix>/NAC/<granularity>/KDK/<key-id>}
\end{center}
Because the KDK is encrypted for each authorized decryptor, the KDK Data packet has additional components:
\begin{center}
	KDK Data Name = \name{/<access manager prefix>/NAC/<granularity>/KDK/<key-id>/ENCRYPTED-BY/<decryptor prefix>/KEY/<decryptor key-id>}
\end{center}
where the key-id is the same as its corresponding KEK, and key identified by decryptor key-id is the decryptor's key that is used to encrypt the KDK.

\subsubsection{Content Key (CK)}

CK name and CK Data packet names follow conventions similar to KDK. The CK Data packet name follows the naming convention:
\begin{center}
	CK Name = \name{/<producer prefix>/CK/<key-id>}

	CK Data Name = \name{/<producer prefix>/CK/<key-id>/ENCRYPTED-BY/<access manager prefix>/NAC/<granularity>/KEK/<key-id>}
\end{center}

\subsection{NAC Scheme}

The main workflow of NAC is as depicted in Fig.~\ref{fig:nac-model}.

\subsubsection{Key Generation and Provision}
The access manager will generate corresponding KDK and KEK pairs as in the access control polices.
It then directly publishes the KEK (in plaintext) Data packets.
The access manager will encrypt the KDK for each authorized decryptor to ascertain that only intended decryptors with necessary access rights and permissions can get the KDK for data under a data prefix specified by the access policies.
Decision on how to grant access is at the sole discretion of the access manager and is outside the scope of NAC design.

When the connection is not stable or the access manager is supposed to go offline after the bootstrapping process for stronger security, the access managers can simply publish KEK and KDKs to in-network data repositories, so that the encryptors and authorized decryptors can continue to work without communicating with the access manager.

\subsubsection{Key Delivery}
KEK and KDK are all named and just like any other NDN Data packet, can directly be fetched through Interest packets carrying the corresponding key names. Both KEK and KDK names can directly convey
\begin{enumerate*} [label=(\roman*)]
	\item who are supposed to use the key
	and
	\item for which set of data the key should be used.
\end{enumerate*}
In this way, encryptors and decryptors can generate the Interests automatically by following the naming convention (Section~\ref{sec:naming}) to fetch KEK and KDK, while the key names allow the decryptors to learn the granularity of the access control.

\subsubsection{Content Encryption}
Encryptors utilize the KEK and symmetric encryption mechanisms like AES-CBC~\cite{RFC3602} to produce encrypted content.
The symmetric encryption key is called CK (content key) in NAC. After fetching the KEK from the access manager or data repositories, an encryptor learns for which granularity the KEK should be used by checking the KEK name. Then it encrypts the data in this granularity with a CK and encrypts the CK with the KEK. The encryptor will wrap the encrypted content with the CK name into a Data packet and the encrypted CK to another Data packet and publish them. Based on the application needs, the CK can also be carried with the ciphertext in one Data packet.

\subsubsection{Content Decryption}
The main purpose of content decryption is for authorized decryptors to use the proper KDK to decrypt the CK and then decrypt the encrypted content. After fetching the content Data packets, a decryptor can learn which CK should be used for decryption. If CK is not with the content, the decryptor fetches the corresponding CK Data packet. After obtaining the CK, the decryptor uses its own KDK to decrypt the CK.
When the decryptor does not have the KDK or the key is outdated, the decryptor can learn the KDK name from the CK Data packet and generate an Interest to fetch the KDK from the network; more details about this can be found in Section~\ref{sec:auto}.

\subsubsection{Access Revocation}
\label{sec:revoke}
In NAC, the access rights are supposed to be short-lived, i.e., to maintain continued access to the content, the access manager needs to periodically update the KEK and KDK pairs.
Such periodic KEK and KDK renewals are transparent to decryptors because decryptors can automatically follow the naming convention to fetch the new KDK when needed instead of periodically querying for the new KDK. The periodic key serves as the baseline of access revocation: when a user is reported to be compromised, the access manager should not grant renewed access rights to the user.
When urgency is the main concern of the access control system, the access manager should send notification to encryptors to use new KEK and generate new CK; the decryptor may also need to re-encrypt all the existing data (i.e., create new versions of the previously created Data packets) with new CK(s) and KEKs to prevent information leakage to and through compromised users.
Note that this does not remove access to any previously published data residing inside in-network caches, as encryptors cannot control state in the distributed system.

\section{NAC based on Attribute-based Encryption}
\label{sec:nac-abe}

We implemented the first prototype of NAC using RSA, but NAC-RSA (or simply referred to as NAC) runs into scalability issues as the number of decryptors increases. In the basic NAC design, access managers directly manage the data access, encrypting KDKs for all authorized decryptors for each granularities. 
For example, assume that there are $n$ soldiers in the battlefield and $m$ authorized granularities. 
To grant each soldier the access rights to $m$ granularities, each user needs to obtain $O(m)$ KDKs, thus the access manager needs to generate $O(m)$ key pairs and produce O($m \times n$) KDK Data packets.
When suffixes are added to achieve fine-granularity, the value $m$ could become much larger as the number of suffix components get added.
For example, the granularity \name{/military/air/aircraftA} contains two sub-granularities \name{/military/air/aircraftA/north} and \name{/military/air/aircraftA/south}.
In this case, the encryptor will create two CKs for the two sub-granularities and encrypt each CK with corresponding KEK, thus decryptors authorized to access the parent granularity need two KDKs to obtain the access rights.

Attribute-based encryption is a type of public-key encryption scheme~\cite{abe}. 
In Ciphertext-Policy Attribute-Based Encryption (CP-ABE)~\cite{cp-abe}, data is encrypted based on an access tree that describes authorized users in terms of attributes, and the users' secret keys are generated over a set of attributes.
CP-ABE makes it possible that the user with the set of attributes which satisfy the encryption attribute policy can decrypt the ciphertext.
As a simple example of CP-ABE, assume a soldier has the attribute set  \{``Soldier'', ``SquadA''\}, the soldier can decrypt the content with the access policy ``Soldier AND SquadA'', but cannot decrypt another ciphertext associated with the policy ``General AND SquadA''.

In this section, we explain how to utilize CP-ABE to realize the NAC and achieve better scalability, and on the other hand, show how NAC scheme can help automate the key delivery in CP-ABE.

\subsection{NAC-ABE with Better Scalability}

We implemented NAC with CP-ABE in our second prototype. In NAC-ABE, \emph{attribute authority} takes responsibility of issuing attributes to the decryptors, and in practice, the attribute authority and access manager can be in the same node. In our battlefield example, the command center plays the additional role of attribute authority.

\begin{figure}[htbp]
	\centering
	\includegraphics[width=0.8\linewidth]{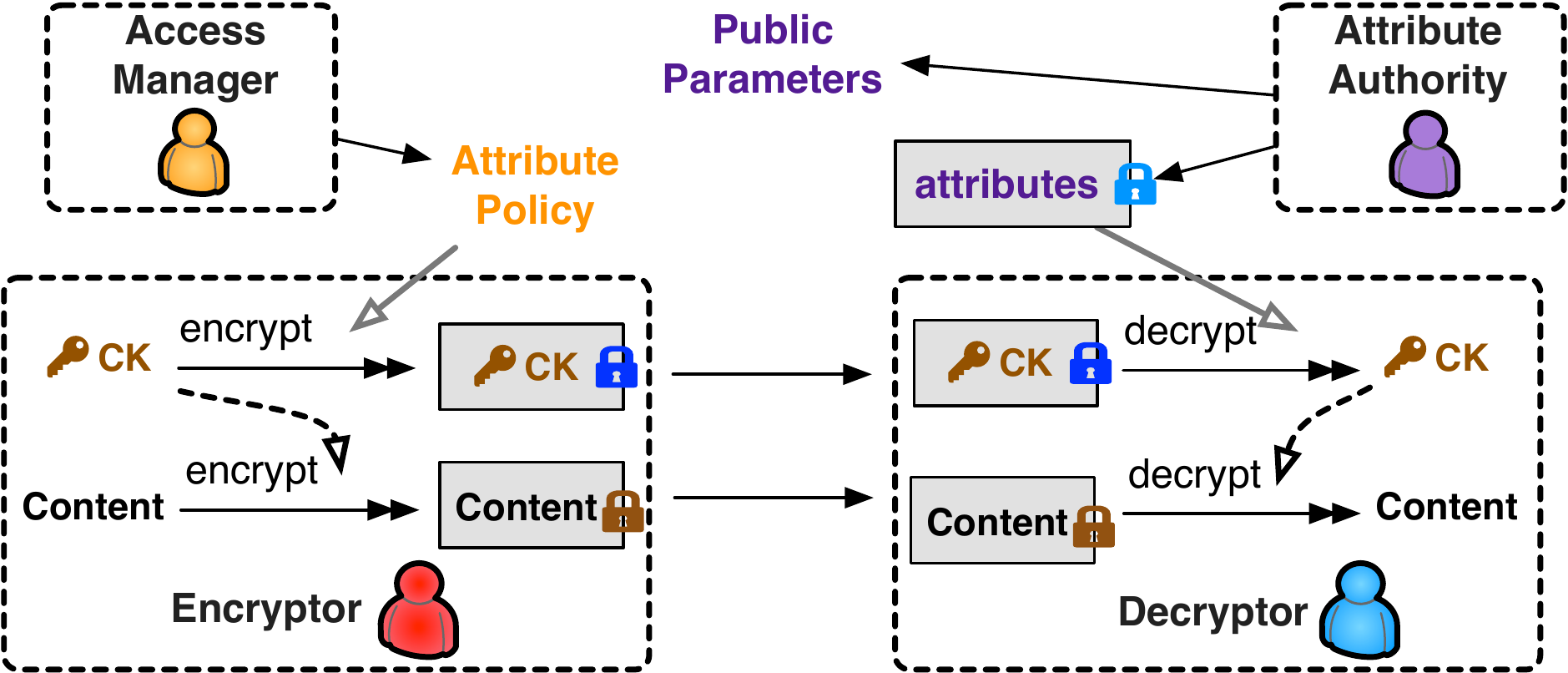}
	\caption{NAC-ABE Scheme}
	\label{fig:abe-model}
\end{figure}

In NAC-ABE, as shown in Fig.~\ref{fig:abe-model}, KEK is an attribute policy while the decryption key is a set of attributes that can satisfy the attribute policy. Different from NAC based on RSA, the encryptor in NAC-ABE encrypts the CK using attribute-based and decryptors decrypt the CK with their attributes.

The attribute authority serves as one level of indirection that allows the system to simply define attributes needed to access their data, and decryptors to have sets of attributes to obtain access rights. After defining the attributes in the system, there is no need to generate decryption keys for each granularity. 
Assuming there are $n$ soldiers, $m$ granularities and $a$ different attributes, the access manager needs to create ABE keys for O($a$) times.
Notably, $a$ is much smaller than $m$: the access controller can combine a small number of attributes with different logic gates (e.g., AND, OR, NOT) to make attribute policies for all the granularities.
Since a decryptor's attributes can be issued in one time, the access manager only needs to generate O($n$) packets.
In practice, this process can be greatly improved by issuing attributes in groups or along with identity certificates.

Confirming to the access revocation design of NAC scheme (Section~\ref{sec:revoke}), in NAC-ABE, attributes issued to decryptors are expected to have limited validity period. 
For example, the attribute "SquadA" may have a name like ``SquadA-July8-2018", indicating that the attributes have effect only for a certain period of time; and based on access manager's requirements, the policy can rotate to a ``fresh" set of attributes.

\subsection{Naming Convention of NAC-ABE}

\begin{figure}[htbp]
	\centering
	\includegraphics[width=0.8\linewidth]{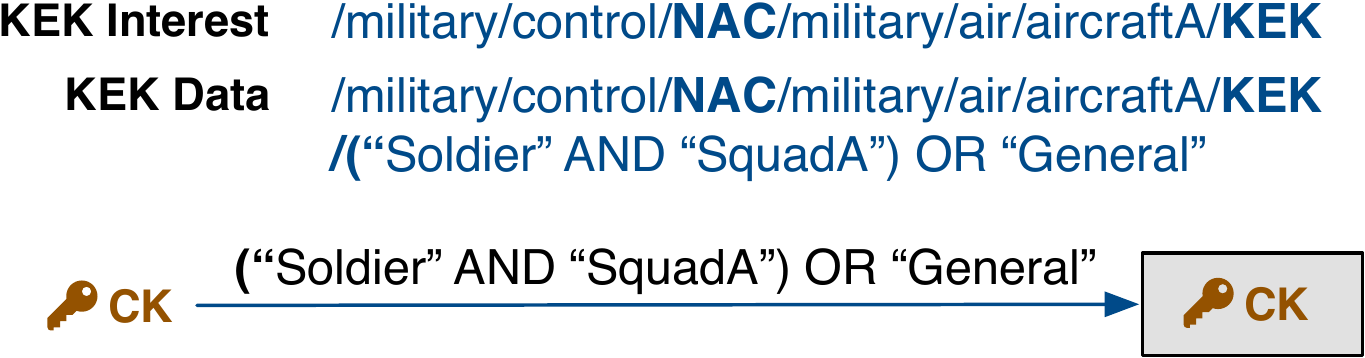}
	\caption{An example of KEK in NAC-ABE}
	\label{fig:abe-kek}
\end{figure}

The naming convention in NAC-ABE follows general NAC conventions described in Section~\ref{sec:naming} with several exceptions. The KEK name follows the same convention but the key-id is no longer a unique identifier but an attribute policy that is used for ABE-based encryption. For example, as shown in Fig.~\ref{fig:abe-kek}, the key-id is ``(Soldier AND SquadA) OR General". Through the name, the decryptor learns which attribute policy should be used to encrypt the data in granularity \name{/military/air/aircraftA}.

When a decryptor needs to fetch an authorized attribute from the attribute authority, the decryptor can generate the attribute Interest packet by following the convention.
\begin{center}
	Attribute Interest Name = \name{/<attribute authority prefix>/ATTRIBUTE/<attribute name>/ENCRYPTED-BY/<decryptor prefix>/KEY/<decryptor key id>}
\end{center}
In NAC-ABE, attributes are expected to be provisioned before the system starts.

\section{Properties of NAC and NAC-ABE}
\label{sec:properties}

\subsection{Automatic Key Delivery}
\label{sec:auto}

\subsubsection{Automatic KEK Retrieval}
Given that access manager's prefix is well-known in an access control system and an encryptor knows\footnote{The knowledge can be configured, defined by a schema, or inferred from the data name. Currently, NAC does not define any specific mechanics for that.} the granularity (name prefix) of its produced data, the encryptor can construct Interest packets for KEK automatically.

For example, assuming aircraft A (Figure~\ref{fig:model}) produces Data packets under the prefix \name{/military/air/aircraftA} and the access manager (i.e., the command center) has the prefix \name{/military/control}, the user can automatically generate the Interest packet by appending the expected data prefix to the access manager's prefix.
The Interest will have the name \name{/military/control/NAC/military/air/aircraftA/KEK}. After sending out the Interest packet, a KEK Data packet with a key identifier called key-id will be fetched.
The key-id is either unique identifier for RSA or an attribute policy string for CP-ABE.

With the naming convention of KEK, there is no need of manual configuration of keys to encryptors, thus improving the usability of the system.
At the same time, named data can be fetched directly by its name, thus there is no need of name services (e.g., DNS) in NAC or NAC-ABE

\subsubsection{Automatic CK and KDK Retrieval}

The naming convention also helps decryptors to collect sufficient keys to finish the decryption.
At the time when the encryptor produces the encrypted content Data packets, the encryptor will explicitly put the CK name into the content.
After getting the content Data packets, the decryptor can extract the CK name from the Data packet and the CK name can directly be used as the Interest packet to fetch the CK Data packets.
Similarly, the fetched CK Data packet name can directly convey the KEK name.
Following the naming convention, a decryptor in NAC can simply flip the \name{KEK} to \name{KDK} and append its decryptor identity to construct an Interest packet for KDK.

\begin{figure}[htbp]
	\centering
	\includegraphics[scale=0.5]{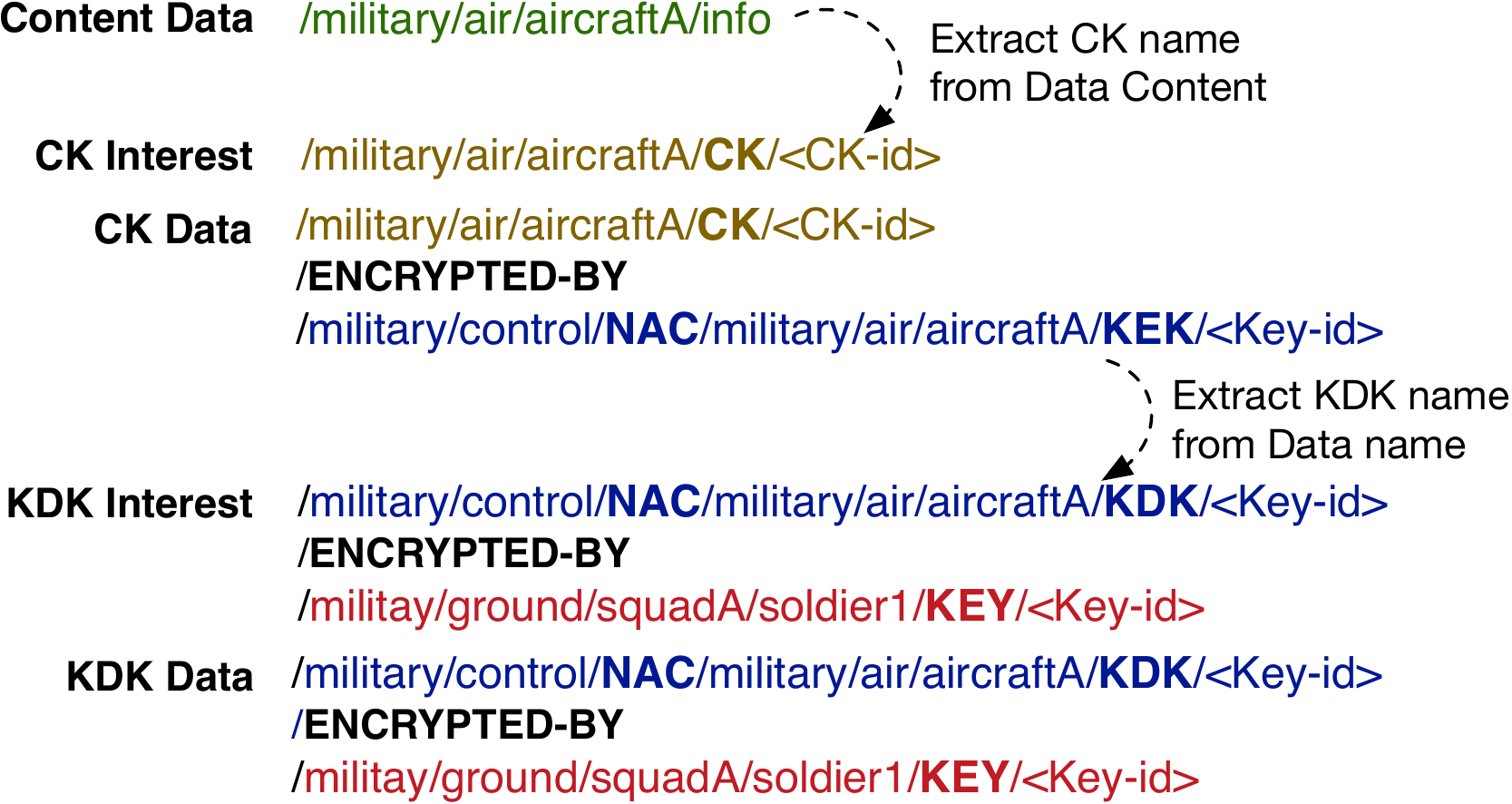}
	\caption{An example of content decryption in NAC}
	\label{fig:nac-naming}
\end{figure}

The naming conventions of CK and KDK allow automation of the whole process of content decryption and related key retrieval.
For example, in NAC, as shown in Figure~\ref{fig:nac-naming}, a decryptor with name \name{/military/ground/squadA/soldier1} wants to decrypt the data \name{/military/air/aircraftA/info} sent from the aircraft A.
By checking the CK field of the content Data packet, the decryptor learns the CK name \name{/military/air/aircraftA/CK/<CK-id>} and uses it to fetch CK Data.
From the name of CK Data packet, the decryptor can directly extract the KEK name \name{/military/control/NAC/military/air/aircraftA/KEK/<Key-id>}.
By changing the component \name{KEK} to \name{KDK} and appending its name \name{/ENCRYPTED-BY/military/ground/squadA/soldier1/KEY/<Key-id>}, the decryptor can send out the KDK Interest and fetch the corresponding KDK that is assigned to this decryptor.

\begin{figure}[htbp]
	\centering
	\includegraphics[scale=0.5]{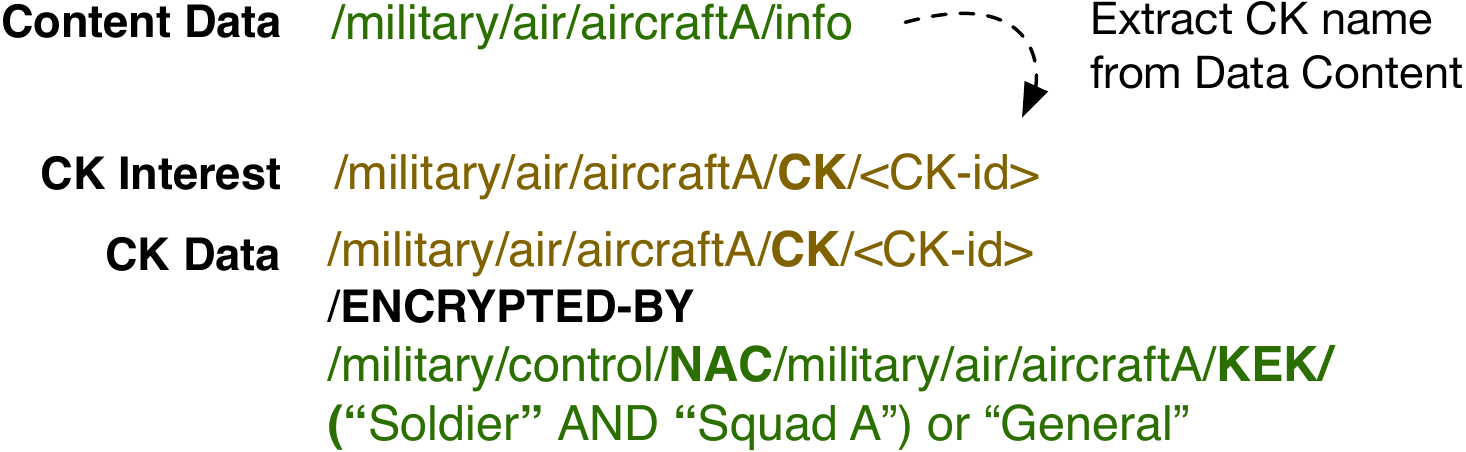}
	\caption{An example of content decryption in NAC-ABE}
	\label{fig:abe-naming}
\end{figure}

Similarly, in NAC-ABE, for example, as shown in Figure~\ref{fig:abe-naming}, a decryptor wants to decrypt the content Data packet \name{/military/air/aircraftA/info}.
The decryptor first fetches the CK back.
As indicated by the CK name, the CK is encrypted by the attribute policy ``(Soldier AND SquadA) or General," which means that soldiers from squad A or the general can access the content.
The soldier then checks whether his existing attributes are sufficient enough, i.e., whether he has attribute ``SquadA" and ``Soldier" or the attribute "General."

\subsection{Fine-Grained Access Control}
\label{sec:granularity}

In NDN, data is named with a structured name.
This allows us to group data with the same properties into the same namespace.
As an illustrative example, Figure~\ref{fig:model} shows the naming prefix for the battlefield application scenario.
Under the prefix \name{/military}, the system allocates a sub-namespace \name{/military/ground} for the data produced by the squads as the ground force; under the \name{/military/ground}, there are three sub-namespaces \name{/military/ground/squad(A/B/C)} representing the data produced by each squad.
Further sub-namespaces can be assigned for finer data production control.
In NAC and NAC-ABE, the access manager will produce KEK with the granularity to be the content prefix and KDK with the decryptor name to be the authorized decryptors.
For example, to grant a user \name{/military/ground/squadA/soldier1} in squad A to access the content produced by a user \name{/military/air/aircraftA}, the access manager can produce the KEK with name \name{/military/control/NAC/military/air/aircraftA/KEK/<key-id1>} and KDK with name \name{/military/control/NAC/military/air/aircraftA/KDK/<key-id1>/ENCRPYPTED-BY/military/ground/squadA/soldier1/KEY/<key-id2>}

In NAC-ABE, besides utilizing structured name, the system can also achieve fine-grained access control by defining attributes based on the granularity needs, enabling the access manager to make attribute policies in a more delicate way.

\begin{figure}[htbp]
	\centering
	\includegraphics[scale=0.45]{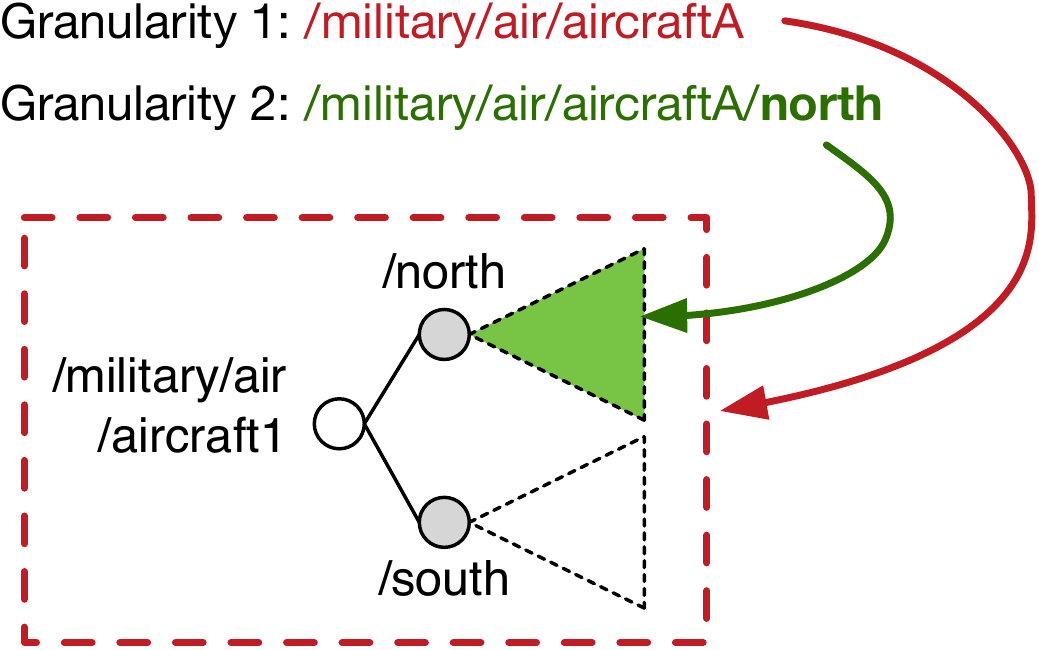}
	\caption{Fine granularity by naming data}
	\label{fig:fine-grained}
\end{figure}

An access manager is able to realize fine-grained access control with NAC and NAC-ABE by restricting the authorized granularity (data name prefix) of the KEK.
In NAC, for example, KEK \name{/military/control/NAC/military/air/aircraft1/KEK/<key-id1>} will be used to encrypt the data produced under \name{/military/air/aircraft1}, as shown in the Figure~\ref{fig:fine-grained}.
By adding the suffix \name{/north} to granularity component of the KEK, the granularity of the KEK is the data produced only in northern battlefield.
Data packet produced under different prefix, e.g., \name{/military/air/aircraft1/south}, cannot be decrypted by the decryptors who are granted the access to \name{/military/air/aircraft1/north}.

\subsection{Support for Intermittent Connectivity}

Different from the channel-based communication model in TCP/IP, NDN's Interest-Data packet exchange and stateful forwarding~\cite{stateful-forwarding} survive when the connectivity is disrupted. In NDN, when a link of the communication path is down, the Data packets can be cached along the path and fetched by future Interests.
Using the scenario shown in Figure~\ref{fig:model}, let us assume that a soldier sent two Interest packets for the encrypted content produced by aircraft A and a KDK from the command center.
If the link between the aircraft gateway and the squad gateway went down before the replied Data packets arrive at the aircraft gateway, two Data packets will still be cached by the aircraft gateway.
Later (after a timeout or when connectivity is restored) the soldier can re-express Interests, which would fetch the content directly from the aircraft gateway instead of from the command center and the aircraft A.

Deploying dedicated large cache storage on the forwarders with intermittent links can greatly improve the data availability.
Importantly, such mechanisms do not require application semantics, instead, they are naturally supported by NDN at the network layer.

\section{Security Assessment}
\label{sec:assessment}

In this paper, we focus on threats that are specific to communication confidentiality and access control and in this section, we explain how NAC mitigates these potential threats.
We also show that some threats including man-in-the-middle (MITM) and denial-of-service (DoS) are natively mitigated by NDN.
We leave threats specific to particular cryptography algorithms, e.g., collusion attack in attribute-based encryption, outside the scope of this paper.

\subsection{Threat Mitigation by NAC}

\paragraph{Eavesdropping}

Attackers may sniff on the broadcast media or retrieve published Data packets from in-network caches.
However, since all the sensitive content (e.g., data, CK, KDK) in NAC are encrypted, even though attackers can collect these Data packets from the broadcast media or cache, the attackers cannot make sense of these ciphertext.

\paragraph{Device Compromise}

Attackers may compromise individual devices to gain the data access that is granted to the unit. There are no means to stop a compromised device from accessing the previously published content, but an access control scheme is supposed to revoke the device's privilege as soon as possible in order to prevent further leak of data. NAC utilizes the short-lived KEK KDK pairs to reduce the information leakage in cases of compromised devices. As mentioned in Section~\ref{sec:revoke}, based on application's needs, the access manager can also initiatively notify encryptors to re-encrypt the content using the new keys before the old keys get expired.

\subsection{Threat Mitigation by NDN}

NAC is based on NDN, and we argue that NDN itself helps in mitigating MITM and DoS attacks.
In NDN, the Data packets directly protected by applications: producers sign each Data packet and consumers verify the signature to ensure the integrity and authenticity.

\paragraph{Man-in-the-Middle Attack}

In NAC, when attackers perform Man-in-the-Middle (MITM) attack and modify the KDK packets, the decryptors will notice the change by verifying the signature.

\paragraph{Denial-of-Service Attack}

Since all the content Data packets and key Data packets are published in the NDN network, these packets can be cached in cache or dedicated data repositories.
When attackers flood the Interest packets for the content or keys, the cache can stop these Interests.
If attackers use forged Interest packets (e.g., append randomness to a valid prefix), mechanism proposed and mentioned in \cite{pap, al2015revisiting, dai2013mitigate,compagno2013poseidon, afanasyev2013interest} can mitigate such attacks.

\section{Performance Evaluation}
\label{sec:evaluation}

We evaluate the performance of NAC scheme in terms of packet efficiency and cryptographic operations. The evaluation is aimed to offer quantitative analysis of our system for NAC users so that the system can fit into their specific hardware and network environment.

\subsection{Bandwidth and Packet Size}
In this section, we set the content size (plaintext) to be 1024 bits and the name of the Data packet is \name{/producer/dataset1/example/data1}.
In NAC, we let the granularity to be \name{/producer/dataset1/example}.
The size value of all the essential Data packets needed in the NAC is shown in Table~\ref{fig:nac-packet-size}.

\begin{table}[htbp]
	\centering
	\caption{NAC Packet Size (Signature Type: SHA256ECDSA)}
	\includegraphics[scale=0.6]{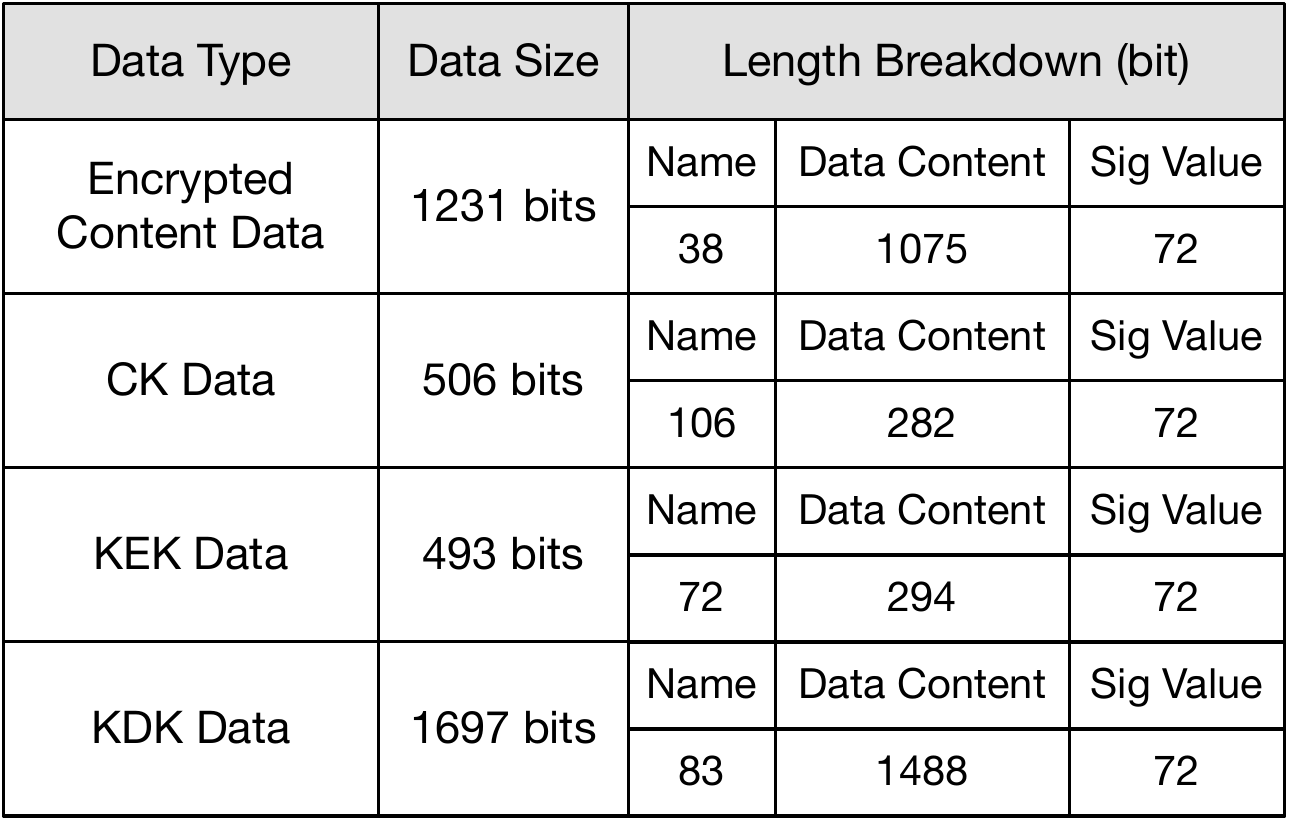}
	\label{fig:nac-packet-size}
\end{table}

We use the same content Data name and granularity in NAC-ABE and have the consumer to obtain 10 attributes: ``attr1" to ``attr10".
The producer encrypts the 1024 bits content with policy ``( attr1 and attr2 ) or attr3".
All the packet size and the breakdown are shown in Table~\ref{fig:nac-abe-packet-size}.

\begin{table}[htbp]
	\centering
	\caption{NAC-ABE Packet Size (Signature Type: SHA256ECDSA)}
	\includegraphics[scale=0.6]{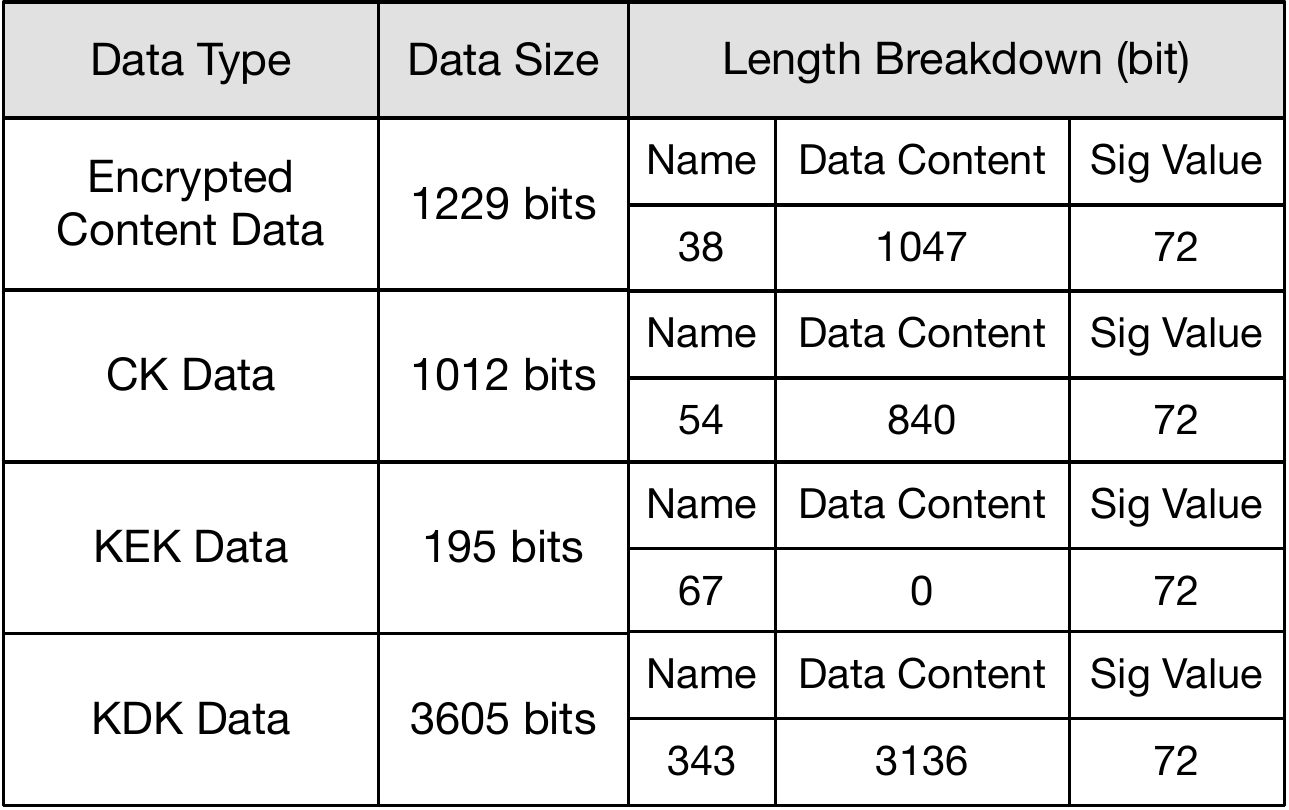}
	\label{fig:nac-abe-packet-size}
\end{table}

\subsection{Cryptographic Operation Evaluation}

Assuming we have $n$ consumer and $m$ granularities in the system, there are totally $x$ content Data packets (each granularity has the same number of Data packets) that will be controlled by the access controller.
In NAC, we let each decryptor has the access right to all $x$ content Data packets.
In NAC-ABE, we assume there are totally $a$ attributes and each decryptor has all the attributes.
The cryptographic operations are listed as Table~\ref{fig:crypto-operation}.

\begin{table}[htbp]
	\centering
	\caption{Cryptographic operations in NAC Scheme}
	\includegraphics[scale=0.6]{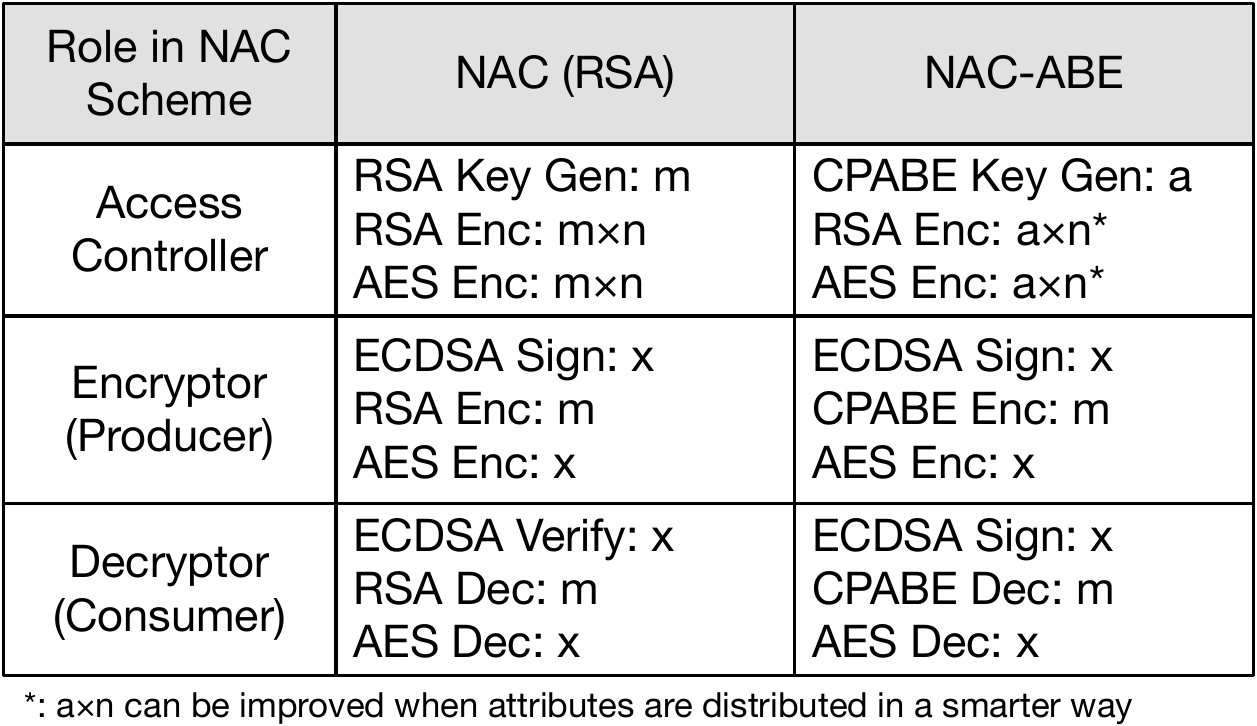}
	\label{fig:crypto-operation}
\end{table}

\subsection{Automated Key Distribution}

In access control system over TCP/IP, to achieve key distribution, the network configuration (e.g., IP address, DNS name) or the equivalent service invocation (e.g., database query) is linear to the number of keys in the system.
In contrast, the network configuration for key delivery in NAC is independent of the number of keys.

\section{Discussion}
\label{sec:disussion}

\subsection{Access Control System over TCP/IP and NDN}

Today's content sharing applications, by and large, rely on a third party to host (e.g., cloud server) their content, and the security in content sharing is provided through encrypted channels like IPsec~\cite{ipsec}, TLS~\cite{tls}, QUIC~\cite{quic}, etc.
However these channels are not directly between content producers and consumers, but between producers and host, and host and consumers.
This practice fails to provide end-to-end confidentiality because it allows a third party, the content host, to see the shared content in plain text, leading to potential privacy concerns and liability on the content hosts.
Furthermore, protected network channels do not directly translate to data confidentiality---data could have been altered before entering the channel and lose confidentiality after it leaves the channel.

To achieve true end-to-end confidentiality and access control, a security system is supposed to decouple the content confidentiality from any hosting party by securing the content directly. More specifically, a content producer should encrypt content at the time of production, then it can control the sharing of its content by managing the distribution of the corresponding decryption keys. In NDN, data is directly protected by the producer at the time of creation, conforming to the idea of data-centric confidentiality.

Regarding the network layer communication, TCP/IP requires both sides of the channel online to setup the communication channel, which does not fit when the connectivity is intermittent and there is a possibility of high packet loss rate, e.g., in battlefield. 
In NDN, the Interest-Data exchange model survives when the connectivity is in poor condition.
Utilizing the in-network caches and dedicated data repositories, NDN provides better data availability compared to connection-based communication.
Therefore, NAC fits more when the underlying network condition is unstable, e.g., a battlefield.

\subsection{Name Confidentiality}
In NDN, data is requested by names but the name itself may reveal sensitive information to some extent. For example, a data name \name{/military/air/aircraftA/north} conveys the information that the content may be related to northern battlefield and produced by an aircraft.
To prevent the information leakage from the data name, the system can hide the Data packet name by obscuring it, e.g., with a hash function.
As pointed by some papers~\cite{ghali2017encryption, ghali2016futility}, even with proper name encryption/hash, the attackers can still infer some information by analyzing the traffic pattern and other characteristics.
However, we argue that such analysis on the traffic is considered to be difficult and time-consuming, and is less harmful compared with unauthorized access and other issues caused by connection-based confidentiality and extra configuration (e.g., which user has which access rights, DNS, IP, etc).

\subsection{Performance and Energy Consumption}
NAC requires the access manager to generate asymmetric key pairs and encryptors/decryptors to perform both symmetric and asymmetric encryption/decryption operations.
These cryptographic operations are considered to be expensive \cite{suo2012security, wang2014performance}, especially for constrained devices, e.g., IoT devices and mobile devices.
NAC does not help in improving the efficiency of the cryptography algorithms but helps to simplify the system realization by automatic key delivery and fine-grained control by names.
Compared to the existing session-based security solution (e.g., IPsec, TLS, QUIC), NAC can be engineered to consume similar amounts of energy.
For example, the same CK can be re-used to encrypt large datasets that are subject to the same access policy.


\section{Conclusion}
\label{sec:conclusion}

Content-based access control model provides a new perspective for end-to-end confidentiality. 
By requiring content encryption at the time of production, the model minimizes the dependency on any intermediate device for access control.
This model naturally fits into the data-centric architecture, such as NDN.
In this paper, we present NAC to provide effective data confidentiality and access control over NDN.


Our work shows that NDN's named data enables NAC to work in a more efficient yet simple way.
\begin{enumerate*} [label=(\roman*)]
	\item The structured namespace of NDN can convey rich contextual information about access control; by defining proper naming conventions for encryption/decryption keys, one conveys access control policies clearly at a fine granularity.
	\item Well-designed naming conventions can significantly facilitate key distribution in the access control system and thus minimize the manual configuration at the network layer.
	\item NDN's data-centric communication model enables NAC to work even with intermittent connectivity.
\end{enumerate*}

\bibliographystyle{IEEEtran}
\bibliography{refs}

\end{document}